\documentclass[article]{IEEEtran}
\onecolumn
\ifCLASSINFOpdf
\else
\fi
\usepackage[cmex10]{amsmath}
\usepackage{amsfonts}
\usepackage{amsmath}
\usepackage{amssymb}
\usepackage{latexsym}

\newtheorem{theorem}{Theorem}
\newtheorem{lemma}{Lemma}
\newtheorem{corollary}{Corollary}

\newtheorem{definition}{Definition}

\begin{document}

\title{Characterization of the Least Periods of the Generalized Self-Shrinking Sequences}
%
%
%

\author{A.~F\'uster-Sabater
\thanks{A.~F\'uster-Sabater is with the Institute of Physical and Information Technologies (ITEFI), (C.S.I.C.), Madrid,
Spain.

 e-mail: amparo@iec.csic.es}
\thanks{Manuscript received September 4, 2013; revised 2013.}}

%
%

\markboth{}%
{Shell \MakeLowercase{\textit{et al.}}: On the Period of the Generalized Self-Shrinking Sequences}
%



\maketitle

\begin{abstract}
In 2004, Y. Hu and G. Xiao introduced the generalized self-shrinking generator, a simple bit-stream generator considered as a specialization of the shrinking generator as well as a generalization of the self-shrinking generator. The authors conjectured that the family of generalized self-shrinking sequences took their least periods in the set $\{1, 2, 2^{L-1}\}$, where $L$ is the length of the Linear Feedback Shift Register included in the generator. In this correspondence, it is proved that the least periods of such generated sequences take values exclusively in such a set. As a straight consequence of this result, other characteristics of such sequences (linear complexity or pseudorandomness) and their potential use in cryptography are also analyzed.
\end{abstract}


%
\IEEEpeerreviewmaketitle

\section{Introduction}

Linear Feedback Shift Registers (LFSRs) \cite{Golomb:1982} are linear
structures currently used in the generation of pseudorandom sequences.
Simplicity, ease of implementation, and good statistical
properties in the output sequences turn LFSRs into natural building blocks for the
design of sequence generators with applications in fields so different as:
spread-spectrum communications, circuit testing, error correcting
codes, numerical simulations or cryptography. Traditionally, pseudorandom sequence generators
involve one or more than one LFSR combined by means of nonlinear functions,
irregular clocking or decimation techniques.

Inside the family of irregularly decimated generators, we can enumerate: a)
the shrinking generator proposed by Coppersmith, Krawczyk and Mansour \cite{Coppersmith:1993}
that includes two LFSRs, b) the self-shrinking generator designed by Meier and
Staffelbach \cite{Meier:1995} involving only one LFSR and c) the generalized self-shrinking
generator proposed by Hu and Xiao \cite{Hu:2004} that can be considered as a specialization
of the shrinking generator as well as a generalization of the self-shrinking generator.
Irregularly decimated generators produce cryptographic sequences
characterized by good correlation features, excellent run distribution,
balancedness, large linear complexity, simplicity of implementation, etc. The underlying idea of
this kind of generators is the irregular decimation of a \textit{m}-sequence according to
the bits of another one. The result of this decimation process is a binary sequence
that will be used as keystream sequence in stream cipher.

This correspondence focuses on the most representative element in this class of irregularly decimated
generators, that is the \textit{generalized self-shrinking generator} \cite{Hu:2004} that generates a family of binary sequences. 
More precisely, the least period of each one of the generalized self-shrinking sequences has been determined.
In this sense, the conjecture formulated by the generalized self-shrinking generator's authors in \cite{Hu:2004} is proved. As a consequence of the previous result, additional characteristics of such sequences have been also considered and analyzed.

The work is organized as follows.
A description of the generalized self-shrinking generator with some fundamental remarks are introduced  in
Section 2. Next in Section 3, a formulation of the generalized self-shrinking sequences based on extended fields is developed. Main results concerning the least periods of such sequences appear
in Section 4. Discussion on cryptographic features of the generalized self-shrinking sequences is given in Section 5.
Finally, conclusions in Section 6 end the paper.

\section{The generalized self-shrinking generator}
This generator can be described as follows:
\begin{itemize}
\item It makes use of two sequences: a \textit{m}-sequence
    $\{a_n\}$ and a shifted version of such a sequence denoted
    by $\{v_n\}$.
\item It relates both sequences by means of a simple decimation rule to generate an output sequence.
\end{itemize}

In mathematical terms, the family of generalized
self-shrinking sequences can be defined as follows \cite{Hu:2004}:
\begin{definition}
Let $\{a_n\}$ be a \textit{m}-sequence over $GF(2)$ with period $2^L - 1$
generated from a maximal-length LFSR of $L$ stages. Let $G$ be a
$L$-dimensional binary vector defined as:
\begin{equation}\label{equation:1}
G = (g_0, g_1, g_2, \ldots, g_{L-1}) \in   GF(2)^L.
\end{equation}
The $n$-th term of the sequence {$v_n$} is computed as:
\begin{equation}\label{equation:2}
v_n = g_0 \,a_n + g_1 \,a_{n+1} +  g_2 \,a_{n+2} + \ldots + g_{L-1} \,a_{n+L-1},
\end{equation}
where the sub-indexes of the sequence $\{a_n\}$ are reduced mod
$2^L - 1$. For $n \geq 0$ the following decimation rule is applied:
\begin{itemize}
\item If $a_n= 1$, then $v_n$ is output.
\item If $a_n= 0$, then $v_n$ is discarded and there is no output bit.
\end{itemize}
In this way, an output sequence $\{b_0, b_1, b_2, \ldots\}$ denoted by
$\{b_n(G)\}$ over $GF(2)$ is generated. Such a sequence is called
a generalized self-shrinking sequence. The sequence family
$B(a_n)=\{\{b_n(G)\}, \,\, G\in GF(2)^L\}$ is called the family of generalized self-shrinking sequences based on the \textit{m}-sequence $\{a_n\}$.
\end{definition}

Recall that $\{a_n\}$ remains fixed while $\{v_n\}$ is the sliding sequence or shifted
version of $\{a_n\}$. The $2^L-1$ nonzero choices of $G$ over $GF(2)^L$ result in the
$2^L -1$ distinct shifts of $\{v_n\}$. For each possible sequence $\{v_n\}$ and after the application of the decimation rule, a new generalized self-shrinking sequence is generated. The sequence family includes the $2^L-1$ generalized self-shrinking sequences plus the identically null sequence corresponding to $G=(0, 0, \ldots, 0)$.

Some important facts extracted from \cite{Hu:2004} are enumerated:
\begin{enumerate}
\item All the sequences in this family are balanced except for sequences $'0000 \ldots 0'$ and $'1111 \ldots 1'$, \cite[Th. 1]{Hu:2004}.
\item By construction, the family of generalized self-shrinking sequences consists of $2^L$ sequences of $2^{L-1}$ bits each of them \cite[Section I]{Hu:2004}. Consequently, the least period of each one of these sequences is a factor of $2^{L-1}$.
\item The family of generalized self-shrinking sequences has structure of Abelian group whose group operation is the bit-wise addition mod 2, the neutral element is the sequence $'0000 \ldots 0'$
and the inverse element of each sequence is the own sequence, \cite[Th. 2]{Hu:2004}.
\item The correlation between two generalized self-shrinking sequences is excellent except for sequences mutually complemented, \cite[Section II]{Hu:2004}.
\item The self-shrinking sequence is a member of the generalized self-shrinking sequence family, \cite[Section I]{Hu:2004}.
\end{enumerate}
Now a distinct representation of the generalized self-shrinking sequences for the study of their least periods is introduced.

\section{\textit{m}-sequences and extended fields}
Let $p(x)$ be the minimal polynomial of the \textit{m}-sequence $\{a_n\}$, that is a primitive polynomial of degree $L$,
\begin{equation}\label{equation:3}
p(x) = p_0 + p_1 x + p_2 x^2 + \ldots + p_L x^L ,
\end{equation}
where $p_i \in   GF(2)$ and $p_0 = p_L =1$. Moreover, if $\alpha$ is a root of $p(x)$, then $\alpha$ is a primitive element in
$GF(2^L)$ the extension field of $GF(2)$ that consists of $0$ and appropriate powers of a primitive element \cite{Lidl:1994}.

Next two well known facts concerning \textit{m}-sequences can be stated \cite{Lidl:1994,Peterson:1972}:
\begin{enumerate}
\item Any arbitrary element, $a_n$, of the \textit{m}-sequence can be written in terms of the \textit{trace function} $Tr_1^L(A\, \alpha)$ as follows:
\begin{equation}\label{equation:4}
a_n = A \; \alpha^{n} + A^{2} \; \alpha^{2n} + \ldots + A^{2^{L-1}} \; \alpha^{2^{L-1} \, n} ,
\end{equation}
where $A \in GF(2^L)$ determines the starting point of the \textit{m}-sequence. At the same time and making use of the \textit{Lth}-order linear recurrence relationship, any term $a_n$ $(0 \leq n < 2^{L}-1)$ can be written as a linear combination of the  first $L$ terms
$(a_0, a_1, a_2, \ldots, a_{L-1})$.

\item Any nonzero element $\alpha^{n}$ $(0 \leq n < 2^{L}-1)$ of $GF(2^L)$ can be uniquely expressed as a linear combination of the elements of the basis $\{1, \alpha, \alpha^2, \alpha^3,\ldots , \alpha^{L-1}\}$.
\end{enumerate}

%
\begin{table}[!t]
\renewcommand{\arraystretch}{1.3}
\caption{\textit{m}-sequence terms and extended field elements}
\label{Table:1}
\centering
\begin{tabular}{|l||l|}
\hline
$\qquad\quad\{a_n\} $ & $\qquad GF(2^L)$\\
\hline
$a_0$ & $1$ \\
\hline
$a_1$ & $\alpha$ \\
\hline
$a_2$ & $\alpha^2$ \\
\hline
$a_3 = a_1 + a_0$ & $\alpha^3 = \alpha + 1$ \\
\hline
$a_4 = a_2 + a_1$ & $\alpha^4 = \alpha^2 + \alpha$ \\
\hline
$a_5 = a_2 + a_1 + a_0$ & $\alpha^5 = \alpha^2 + \alpha + 1$ \\
\hline
$a_6 = a_2 + a_0$ & $\alpha^6 = \alpha^2 + 1$ \\
\hline
\end{tabular}
\end{table}
Via the minimal polynomial, there is a one-to-one correspondence $a_n \rightarrow \alpha^{n}$ $(n = 0, 1, \ldots, 2^L - 2)$ between the \textit{nth}-element, $a_n$, of the \textit{m}-sequence expressed in terms of $(a_0, a_1, a_2, \ldots, a_{L-1})$ and the \textit{nth}-power $\alpha^{n}$ written in terms of the basis $\{1, \alpha, \alpha^2, \ldots , \alpha^{L-1}\}$. Table \ref{Table:1} shows such a correspondence for $L = 3$ and $p(x) = x^3 + x + 1$.

As this work focuses on the period of the generalized self-shrinking sequence family, throughout the paper and without loss of generality we consider that the first $L$ bits of $\{a_n\}$ are $(a_0, a_1, \ldots, a_{L-2}, a_{L-1}) = (0, 0, \ldots, 0, 1)$. Thus, just the terms $a_k$ including $a_{L-1}$ in their linear decomposition satisfy $a_k = 1$. In the same way, just the powers $\alpha^k$ corresponding to previous $a_k$ include $\alpha^{L-1}$ in their linear decomposition.

\begin{table}[!t]
\renewcommand{\arraystretch}{1.3}
\caption{Generation of a generalized self-shrinking sequence}
\label{Table:2}
\centering
\begin{tabular}{|l||l||l||l|}
\hline
$\,\,\{a_n\} $ & $ Mapping$ & $\;\;\;\{v_n\}$ & $\;\;\;\{b_n(\alpha^4)\}$\\
\hline
$a_0 = 0$ & $1 \rightarrow \alpha^{4}$ & $v_0 = a_4$ & \\
\hline
$a_1 = 0$ & $\alpha \rightarrow \alpha^{5}$ & $v_1 = a_5$ & \\
\hline
$a_2 = \textbf{1}$ & $\alpha^2 \rightarrow \alpha^{6}$ & $v_2 = a_6$ & $b_0 = a_6 = 1$\\
\hline
$a_3 = 0$ & $\alpha^3 \rightarrow 1$ & $v_3 = a_0$ & \\
\hline
$a_4 = \textbf{1}$ & $\alpha^4 \rightarrow \alpha$ & $v_4 = a_1$ & $b_1 = a_1 = 0$\\
\hline
$a_5 = \textbf{1}$ & $\alpha^5 \rightarrow \alpha^{2}$ & $v_5 = a_2$ & $b_2 = a_2 = 1$\\
\hline
$a_6 = \textbf{1}$ & $\alpha^6 \rightarrow \alpha^{3}$ & $v_6 = a_3$ & $b_3 = a_3 = 0$\\
\hline

\end{tabular}
\end{table}
Since the \textit{m}-sequence terms $\{a_0, a_1, a_2, a_3, \ldots\}$ are associated with the nonzero elements $\{1, \alpha, \alpha^2, \alpha^3,\ldots \}$ of $GF(2^L)$, respectively, then the terms $\{v_0, v_1, v_2, v_3, \ldots\}$ of the sliding sequence are associated with the nonzero elements $\{\alpha^s, \alpha^{(s+1)}, \alpha^{(s+2)}, \alpha^{(s+3)}, \ldots \}$ of $GF(2^L)$, respectively. Thus, there are $2^L - 1$ mappings:
\begin{equation}\label{equation:5}
1 \rightarrow \alpha^{s} \qquad (s = 0, 1, \ldots, 2^L - 2)
\end{equation}
that denote the possible shifts of the sliding sequence $\{v_n\}$ regarding the \textit{m}-sequence $\{a_n\}$. In addition, the mapping $1 \rightarrow \alpha^{s}$ univocally determines the correspondence among the other elements of the extended field, that is: $1 \rightarrow \alpha^{s}$ implies $\alpha \rightarrow \alpha^{s+1}$, $\alpha^2 \rightarrow \alpha^{s+2}$, $\alpha^3 \rightarrow \alpha^{s+3}$,$\ldots$ and so on. Once the mapping has been defined, the application of the decimation rule allows the generation of the corresponding generalized self-shrinking sequence now denoted by $\{b_n(\alpha^s)\}$. Table \ref{Table:2} shows $1 \rightarrow \alpha^{4}$ and $\{b_n(\alpha^4)\}$ for the same $L$ and $p(x)$ as before.

The mapping $1 \rightarrow 1$ generates the generalized self-shrinking sequence identically 1, $'1111 \ldots 1'$, as in this case $\{a_n\}$ and $\{v_n\}$ coincide.

The mapping $1 \rightarrow 0$ generates the generalized self-shrinking sequence identically null, $'0000 \ldots 0'$, as in this case $\{v_n\}$ is the identically null sequence too.

Both sequences, $'1111 \ldots 1'$ and $'0000 \ldots 0'$, are elements in the Abelian group of generalized self-shrinking sequences with period $T = 1$.

Now other mappings generating sequences with greater period will be considered.

\begin{lemma}\label{lemma:1}
Let $\alpha$ be a primitive element in $GF(2^L)$. Then, there is a unique integer $p$, $(L-1 \leq p < 2^L - 2)$, such that $\alpha^p$ and $\alpha^{p+1}$ two consecutive powers of $\alpha$ in $GF(2^L)$ satisfy:
\begin{equation}\label{equation:6}
\alpha^{p+1} = \alpha^{p} + 1.
\end{equation}
\end{lemma}
\textit{Proof:} Denote $\alpha^{m} = \alpha +1$. If $\alpha^{p}$ is the unique multiplicative inverse of  $\alpha^{m}$ that is $\alpha^{m} \cdot \alpha^{p} = 1$ with $p = (2^L - 1) - m$, then
\begin{equation}\label{equation:7}
(\alpha +1) \cdot \alpha^{p} = \alpha^{p+1} + \alpha^{p} = 1.
\end{equation}
\hfill $\Box$

From the previous lemma, the following theorem can be formulated.
\begin{theorem}\label{theorem:1}
The mappings $1 \rightarrow \alpha^{p+1}$ and $1 \rightarrow \alpha^{p}$ with $p$ defined as before generate the generalized self-shrinking sequences with period $T = 2$. 
\end{theorem}
\textit{Proof:} Let $a_{n_i}$ $(0 \leq i < 2^{L-1})$ be the $2^{L-1}$ terms of $\{a_n\}$ such that $a_{n_i} = 1$, thus the corresponding power $\alpha^{n_i}$ includes $\alpha^{L-1}$ in its linear decomposition.
Writing the mapping $1 \rightarrow \alpha^{p+1}$, we have:
\[
\begin{tabular}{l}
$1 \;\,\,\rightarrow \alpha^{p} + 1$\\
$\alpha \,\;\rightarrow \alpha^{p} + 1 + \alpha$\\
$\alpha^2 \rightarrow \alpha^{p} + 1 + \alpha + \alpha^2$\\
\;\vdots \qquad \qquad \quad \vdots\\
$\alpha^{n_0} \rightarrow \alpha^{p} + 1 + \alpha + \ldots + \alpha^{n_0}$\\
\;\vdots \qquad \qquad \quad \vdots \qquad \qquad \quad \vdots\\
$\alpha^{n_1} \rightarrow \alpha^{p} + 1 + \alpha + \ldots + \alpha^{n_0} + \ldots + \alpha^{n_1}$\\
\;\vdots \qquad \qquad \quad \vdots \qquad \qquad \quad \vdots \qquad \qquad \quad \vdots\\
$\alpha^{n_2} \rightarrow \alpha^{p} + 1 + \alpha + \ldots + \alpha^{n_0} + \ldots + \alpha^{n_1} + \ldots + \alpha^{n_2}$\\
\;\vdots \\
        \end{tabular}
\]
It can be noticed that if $\alpha^{p} + 1 + \alpha + \ldots + \alpha^{n_0}$ contains the power $\alpha^{L-1}$ an even (odd) number of times, then
$\alpha^{p} + 1 + \alpha + \ldots + \alpha^{n_0} + \ldots + \alpha^{n_1}$ contains the power $\alpha^{L-1}$ an odd (even) number of times and
$\alpha^{p} + 1 + \alpha + \ldots + \alpha^{n_0} + \ldots + \alpha^{n_1} + \ldots + \alpha^{n_2}$ contains the power $\alpha^{L-1}$ an even (odd) number of times , $\ldots$ and so on. Consequently,
the terms of the sliding sequence $v_{n_0}, v_{n_1}, v_{n_2}, \ldots $ contain alternatively an even, odd, even, $\ldots$ (odd, even, odd, $\ldots$) number of $1's$ in their linear decompositions, respectively. Therefore, the corresponding generalized self-shrinking sequence will be $\{b_n(\alpha^{p+1})\} = \{u, \bar{u}, u, \bar{u}, \ldots\}$ with $u, \bar{u} \in GF(2)$, $u$ being an arbitrary bit and $\bar{u}$ the complemented bit.
Analogous reasoning follows for the mapping  $1 \rightarrow \alpha^{p}$.
\hfill $\Box$

\begin{corollary}\label{corollary:1}
Provided that the \textit{m}-sequence $\{a_n\}$ starts at the initial state $(a_0, a_1, \ldots, a_{L-2}, a_{L-1}) = (0, 0, \ldots, 0, 1)$, the mappings
$1 \rightarrow \alpha^{p+1}$ and $1 \rightarrow \alpha^{p}$ generate the generalized self-shrinking sequences $'0101 \ldots 01'$ and $'1010 \ldots 10'$, respectively.
\end{corollary}
\textit{Proof:}
According to (\ref{equation:6}), $\alpha^{p}$ and $\alpha^{p+1}$ can be written as:
\begin{equation}\label{equation:8}
\alpha^{p} = c_{L-1}\alpha^{L-1} + c_{L-2}\alpha^{L-2} + \ldots + c_1 \alpha + c_0 \alpha^0
\end{equation}
\begin{equation}\label{equation:9}
\alpha^{p+1} = c_{L-1}\alpha^{L-1} + c_{L-2}\alpha^{L-2} + \ldots + c_1 \alpha + \bar{c}_0 \alpha^0,
\end{equation}
with $c_j \in GF(2)$.

If $c_{L-1}=0$ and $i$ the greatest index for which $c_i\neq 0$, then
\[
\begin{tabular}{l}
$\alpha^{p} = \qquad c_{i}\alpha^{i} + c_{i-1}\alpha^{i-1} + \ldots + c_1 \alpha + c_0 \alpha^0$ \\

$\alpha^{p+1} = c_{i}\alpha^{i+1} + c_{i-1}\alpha^{i} \;\;\; \,+  \ldots + c_1 \alpha^2 + c_0 \alpha$,
\end{tabular}
\]
and (\ref{equation:6}) would not hold anymore. Thus, $c_{L-1} \neq 0$ and $\alpha^{p}$ contains the power $\alpha^{L-1}$. Therefore,
\begin{itemize}
  \item For the mapping $1 \rightarrow \alpha^{p+1}$,

  $\alpha^{p} + 1 + \alpha + \ldots + \alpha^{n_0}$  implies $v_{n_0} = 0 \qquad$  (even number of terms $\alpha^{L-1}$)

  $\alpha^{p} + 1 + \alpha + \ldots + \alpha^{n_0} + \ldots + \alpha^{n_1}$  implies $v_{n_1} = 1 \qquad$ (odd number of terms $\alpha^{L-1}$)

  $\alpha^{p} + 1 + \alpha + \ldots + \alpha^{n_0} + \ldots + \alpha^{n_1} + \ldots + \alpha^{n_2}$  implies $v_{n_2} = 0 \qquad$  (even number of terms $\alpha^{L-1}$)

  $\ldots$ and so on. Thus, the generalized self-shrinking sequence is $'0101 \ldots 01'$.
  \item For the mapping $1 \rightarrow \alpha^{p}$, the reasoning is analogous giving rise to the generalized self-shrinking sequence $'1010 \ldots 10'$.
\hfill $\Box$
\end{itemize}
\section{Main results}
We have seen that the generalized self-shrinking sequences with $T > 1$ are balanced. Nevertheless, a stronger condition related to balancedness of specific subsequences can be formulated.

\begin{theorem}\label{theorem:2}
Let $\{b_0, b_1, b_2, b_3, \ldots\}$ be a generalized self-shrinking sequence with $T > 2$. Then the subsequences $\{b_{2i}\}$ and $\{b_{2i + 1}\}$ $(i = 0, 1, 2, \ldots )$ are balanced too.
\end{theorem}
\textit{Proof:} It can be proved by contradiction. If the subsequence $\{b_{2i}\}$ had a number of $0's$ greater than the number of $1's$, then $\{b_{2i + 1}\}$ would have a number of $1's$ greater than the number of $0's$ in order to guarantee the balancedness of $\{b_n\}$. Thus, bit-wise adding $\{b_n\}$ and $'0101 \ldots 01'$ the resulting sequence would have more $0's$ than $1's$ and would not be balanced. Nevertheless, the resulting sequence is an element of the generalized self-shrinking sequence Abelian group and it has to be balanced. Conversely, if the subsequence $\{b_{2i}\}$ had a number of $1's$ greater than the number of $0's$, then $\{b_{2i + 1}\}$ would have more $0's$ than $1's$. Thus, bit-wise adding $\{b_n\}$ and $'0101 \ldots 01'$ the resulting sequence would have more $1's$ than $0's$ and would not be an element of the generalized self-shrinking sequence Abelian group.

\hfill $\Box$

According to Theorem \ref{theorem:2}, a generalized self-shrinking sequence with e.g., $T = 4$, would be of the form $\{u_0, u_1, \bar{u}_0, \bar{u}_1, \ldots\}$ with $u_0, u_1, \bar{u}_0, \bar{u}_1 \in GF(2)$, $u_0, u_1$ being arbitrary bits and $\bar{u}_0, \bar{u}_1$ the complemented bits. Thus,  such a sequence would be the interleaving of two sequences $\{u_0, \bar{u}_0, u_0, \bar{u}_0, \ldots\}$ and $\{u_1, \bar{u}_1, u_1, \bar{u}_1, \ldots\}$. Therefore, a \textit{block} (run of 1's in Golomb's terminology \cite{Golomb:1982}) of length $l > 2$ in the \textit{m}-sequence $\{a_n\}$ would correspond to a succession of $l$ bits in the sliding sequence $\{v_n\}$ satisfying $v_n = \bar{v}_{n+2}$.

The generalization of this idea gives rise to the following results.

\begin{lemma}\label{lemma:2}
For each integer $s$, $(0 < s < 2^{L}-1)$, there exists a unique integer $d$, $(0 < d < 2^{L}-1)$, such that the pair  $(\alpha^{s+d}, \alpha^{s})$ in $GF(2^L)$ satisfy:
\begin{equation}\label{equation:10}
\alpha^{s+d} = \alpha^{s} + 1.
\end{equation}
\end{lemma}

\textit{Proof:} Since $\alpha^{s}$ is an element of $GF(2^L)$, then it has a unique multiplicative inverse $\alpha^{m_s}$ such that $\alpha^{s} \cdot \alpha^{m_s} = 1$ where $m_s = (2^L - 1) - s$. Denote  $\alpha^{d} = \alpha^{m_s} + 1$. Thus,
\begin{equation}\label{equation:11}
\alpha^{s} \cdot \alpha^{m_s} = \alpha^{s} \cdot (\alpha^{d} + 1) = \alpha^{s+d} + \alpha^{s} = 1.
\end{equation}
\hfill $\Box$

Thus, the elements of $GF(2^L)$ (except for $0$ and $1$) can be grouped in $2^{L-1} - 1$ pairs $(\alpha^{s+d}, \alpha^{s})$ satisfying the following property.
\begin{lemma}\label{lemma:3}
The mappings $1 \rightarrow \alpha^{s+d}$ and $1 \rightarrow \alpha^{s}$ with $s$ and $d$ defined as in lemma \ref{lemma:2} generate complemented generalized self-shrinking sequences.
\end{lemma}
\textit{Proof:} It follows trivially that the mapping $1 \rightarrow \alpha^{s+d}= \alpha^{s} + 1$ is a linear mapping as so is the operation $+$ in $GF(2^L)$. Thus, according to (\ref{equation:11}) the generalized self-shrinking sequences corresponding to $1 \rightarrow \alpha^{s+d}$ and $1 \rightarrow \alpha^{s}$ satisfy:
\begin{equation}\label{equation:12}
\{b_n(\alpha^{s+d})\} = \{b_n(\alpha^{s})\} + \{b_n(1)\}.
\end{equation}
The fact that the mapping $1 \rightarrow 1$ generates the generalized self-shrinking sequence identically $1$ completes the proof.
\hfill $\Box$

From the previous lemmas the following theorem can be formulated.
\begin{theorem}\label{theorem:3}
Let $\{a_n\}$ be a \textit{m}-sequence, $\{v_n\}$ the sliding sequence corresponding to the mapping $1 \rightarrow \alpha^{s}$ and $d$ the integer associated with $s$ in lemma 2. Then the following statements hold:
\begin{itemize}
\item If $a_n= 1$, then $v_n = \overline{v}_{n+d}$.
\item If $a_n= 0$, then $v_n = v_{n+d}$.
\end{itemize}
\end{theorem}
\textit{Proof:} If $\{v_n\}$ and $\{v_{n+d}\}$ are the sliding sequences corresponding to the mappings $1 \rightarrow \alpha^{s}$ and $1 \rightarrow \alpha^{s+d}$, respectively, ($\{v_{n+d}\}$ is the sliding sequence shifted $d$ positions regarding $\{v_{n}\}$) then the bit-wise addition of both sequences gives rise to a shifted version of the same \textit{m}-sequence denoted by $\{z_n\}$, $\{z_n\} = \{v_n\} + \{v_{n+d}\}$ $(n\geq 0)$, with $2^{L-1}$ $1's$ and $2^{L-1} - 1$ $0's$. According to lemma \ref{lemma:3} the generalized self-shrinking sequences $\{b_n(\alpha^{s})\}$ and $\{b_n(\alpha^{s+d})\}$ are complemented, then
\begin{itemize}
\item The $1's$ of the sequence $\{z_n\}$ correspond to the terms of $\{v_n\}$ and $\{v_{n+d}\}$ that are distinct as they come from the addition of both generalized self-shrinking sequences. These bits are associated with $a_n= 1$.
\item The $0's$ of the sequence $\{z_n\}$ correspond to the terms of $\{v_n\}$ and $\{v_{n+d}\}$ that are equal and that not appear in the generalized self-shrinking sequences. These bits are  associated with $a_n= 0$.
\end{itemize}
\hfill $\Box$

\begin{table}[!t]
\renewcommand{\arraystretch}{1.3}
\caption{Pairs $(s, d)$ and their corresponding $\{v_n\}$ sequences}
\label{Table:3}
\centering
\begin{tabular}{|l||c||c||c||c||c|}
\hline
$s$ & $$ & $3$ & $7$ & $9$ & $14$\\
\hline \hline
$$ & $\{a_n\}$ & $\{v_n\}_3$ & $\{v_n\}_7$ & $\{v_n\}_{9}$ & $\{v_n\}_{14}$\\
\hline
$$ & $0$ & $1$ & $1$ & $1$ & $1$ \\
\hline
$$ & $0$ & $0$ & $0$ & $0$ & $0$ \\
\hline
$$ & $0$ & $0$ & $1$ & $1$ & $0$ \\
\hline
$$ & $\textbf{1}$ & $\textbf{1}$ & $\textbf{0}$ & $\textbf{1}$ & $\textbf{0}$ \\
\hline
$$ & $0$ & $1$ & $1$ & $1$ & $1$ \\
\hline
$$ & $0$ & $0$ & $1$ & $1$ & $0$ \\
\hline
$$ & $\textbf{1}$ & $\textbf{1}$ & $\textbf{1}$ & $\textbf{0}$ & $\textbf{0}$\\
\hline
$$ & $\textbf{1}$ & $\textbf{0}$ & $\textbf{1}$ & $\textbf{0}$ & $\textbf{1}$ \\
\hline
$$ & $0$ & $1$ & $0$ & $0$ & $1$ \\
\hline
$$ & $\textbf{1}$ & $\textbf{1}$ & $\textbf{0}$ & $\textbf{1}$ & $\textbf{0}$ \\
\hline
$$ & $0$ & $1$ & $0$ & $0$ & $1$ \\
\hline
$$ & $\textbf{1}$ & $\textbf{1}$ & $\textbf{1}$ & $\textbf{0}$ & $\textbf{0}$ \\
\hline
$$ & $\textbf{1}$ & $\textbf{0}$ & $\textbf{0}$ & $\textbf{1}$ & $\textbf{1}$ \\
\hline
$$ & $\textbf{1}$ & $\textbf{0}$ & $\textbf{0}$ & $\textbf{1}$ & $\textbf{1}$ \\
\hline
$$ & $\textbf{1}$ & $\textbf{0}$ & $\textbf{1}$ & $\textbf{0}$ & $\textbf{1}$ \\
\hline \hline
$d$ & $$ & $11$ & $2$ & $13$ & $4$\\
\hline

\end{tabular}
\end{table}

Table \ref{Table:3} shows different sliding sequences $\{v_n\}$ corresponding to $(s, d)$ for $L = 4$ and $p(x) = x^4 + x + 1$. First and last row depict different values of $s$ and $d$, respectively. Moreover, in order to clarify the notation $\{v_n\}_s$ denotes the sliding sequence associated with the mapping $1 \rightarrow \alpha^{s}$. For the pair $(s, d) = (7, 2)$, $\{v_n\}_7$ denotes the sliding sequence
shifted $7$ positions regarding the \textit{m}-sequence $\{a_n\}$. The bits of $\{v_n\}_7$ in bold correspond to the $1's$ of $\{a_n\}$ as well as they satisfy the equality $v_n = \overline{v}_{n + 2}$. The remaining bits of $\{v_n\}_7$ correspond to the $0's$ of $\{a_n\}$ as well as they satisfy the equality $v_n = v_{n + 2}$.

Recall that for $(s, d) = (7, 2)$ the sequences $\{b_n(\alpha^{7})\}$ and $\{b_n(\alpha^{7+2})\}$ are complemented as well as for $(s, d) = (3, 11)$ the sequences $\{b_n(\alpha^{3})\}$ and $\{b_n(\alpha^{3+11})\}$ are complemented too.

The previous results allow us to formulate the main theorem:
\begin{theorem}\label{theorem:4}
The elements of the generalized self-shrinking sequence family do not have least periods of the form $T = 2^{j}$ with $(j = 2, 3, \ldots, L-2)$ where $L$ is the number of stages in the maximal-length LFSR.
\end{theorem}
\textit{Proof:} According to theorem \ref{theorem:3}, the hypothetical generalized self-shrinking sequences with least period $T = 2^{j}$ with $(j = 2, 3, \ldots, L-2)$ should be of the form:
\begin{equation}\label{equation:13}
\{u_0, u_1, u_2\ldots, u_{d-1}, \overline{u}_0, \overline{u}_1, \overline{u}_2, \ldots, \overline{u}_{d-1}\}
\end{equation}
with $u_i, \bar{u}_i \in GF(2)$ and least period $T = 2 \cdot d$, $d$ being a power of 2. That is, this bit configuration in (\ref{equation:13}) would correspond to the interleaving of $d$ sequences $\{u_i, \overline{u}_i, u_i, \overline{u}_i, \ldots \}$ $(i = 0, 1, \ldots, d-1)$ in such a way that the terms $u_i$ and $\overline{u}_i$ in the interleaved sequence would be separated $d$ positions. Although there are other possible least periods different from (\ref{equation:13}) satisfying the conditions of theorem \ref{theorem:2}, nevertheless there exists a unique mapping $1 \rightarrow \alpha^{s}$ with all terms $u_i, \overline{u}_i$ separated $d$ positions.

On the other hand, the number of \textit{blocks} and \textit{gaps} of any length in a \textit{m}-sequence is perfectly quantified \cite{Golomb:1982}. Moreover, the blocks of $\{a_n\}$ output bits in the generalized self-shrinking sequences while the gaps do not. Thus, for $d = 2^k$ $(k = 1, \ldots, L-3)$ the gaps of length $l = \dot{d}$ preserve the periodicity of the bits in (\ref{equation:13}) whereas the gaps of length $l \neq  \dot{d}$ break such a periodicity. As the gap length $l$ ranges in the interval $[1, 2, 3, \ldots, L-1]$, then there will always be gaps of length $l \neq  \dot{d}$ that prevent the generalized self-shrinking sequences from being periodic with $T = 2^{j}$ $(j = 2, 3, \ldots, L-2)$.

For $d = 1$, any gap of length $l$ satisfies $l = \dot{1}$ therefore there exist generalized self-shrinking sequences with $T = 2 \cdot 1$ as proved in theorem \ref{theorem:1}.
\hfill $\Box$

Now the specific least periods of the generalized self-shrinking sequences can be stated.
\begin{corollary}\label{corollary:2}
The family of generalized self-shrinking sequences takes their least periods in the set $\{1, 2, 2^{L-1}\}$. In fact,

$T = 1$ for the mappings $1 \rightarrow 0$ and $1 \rightarrow 1$.

$T = 2$ for the mappings $1 \rightarrow \alpha^{p+1}$ and $1 \rightarrow \alpha^{p}$.

$T = 2^{L-1}$ for the remaining mappings.
\end{corollary}

\textit{Proof:} The result is a straight consequence of the previous theorems.
\hfill $\Box$

The least period of the self-shrinking sequence, the output sequence of the self-shrinking generator as well as an element of this family, deserves particular attention.

\begin{lemma}\label{lemma:4}
The mapping $1 \rightarrow \alpha^{2^{L-1}}$ generates the self-shrinking sequence.
\end{lemma}
\textit{Proof:} Let $\{z_n\}$ be a \textit{m}-sequence with two subsequences $\{z_{2n}\} = \{a_{n}\}$ and $\{z_{2n+1}\} = \{v_{n}\}$ $(n \geq 0)$. By construction, the self-shrinking generator compares the previous sequences $\{a_{n}\}$ and $\{v_{n}\}$ to generate the self-shrinking sequence. Then,
\begin{eqnarray}\label{equation:14}
a_{n+2^{L-1}} = z_{2(n+2^{L-1})} = z_{2n+2^L} & = & \nonumber \\
z_{2n+1+2^L -1} = z_{2n+1} & = & v_n .
\end{eqnarray}

Thus, the shift of the sliding sequence $\{v_{n}\}$ regarding the \textit{m}-sequence $\{a_{n}\}$ equals $s = 2^{L-1}$.
\hfill $\Box$

\begin{corollary}\label{corollary:3}
Except for the case $p = 2^{L-1}$, the least period of the self-shrinking sequence $T_{ss}$ is
\begin{equation}\label{equation:15}
T_{ss} = 2^{L-1}.
\end{equation}
\end{corollary}
\textit{Proof:} The result is a straight consequence of corollary \ref{corollary:2} and lemma \ref{lemma:4}.
\hfill $\Box$

Remark that the lower bound for the least period of the self-shrinking sequence given by Meier and Staffelbach in \cite{Meier:1995}
\begin{equation}\label{equation:16}
T_{ss} \geq 2^{\lfloor L/2 \rfloor},
\end{equation}
is much less than the value given in (\ref{equation:15}).

At the same time and for $L = 3$ and $p(x) = x^3 + x + 1$, the integer $p$ happens to be $p = 2^{L-1} = 4$. Hence, in this case the self-shrinking sequence $\{b_n(\alpha^4)\}$ equals the sequence $'0101 \ldots 01'$ with least period $T_{ss} = 2 < 2^{2}$ what justifies the results found in \cite[Table 1]{Meier:1995}.

\section{Further characteristics of the generalized self-shrinking sequences}
Apart from least periods other cryptographic parameters can be considered for this family of sequences.

\subsection{Linear Complexity of the generalized self-shrinking sequences}
The linear complexity $(LC)$ of a generated sequence is a very used indicator of the security of a stream cipher, see \cite{Menezes:2001, Paar:2010}. As the least periods of the generalized self-shrinking sequences are factors of $2^{L-1}$, then the linear complexity of such sequences can be quantified as follows.
\begin{theorem}\label{theorem:5}
The linear complexity of the generalized self-shrinking sequences with least period $T = 2^{L-1}$ satisfies
\begin{equation}\label{equation:17}
2^{L-2} < LC < 2^{L-1}.
\end{equation}
\end{theorem}
\textit{Proof:} The lower bound follows from the fact \cite{Hu:2004} that the linear complexity of a periodic sequence is greater than $2^{k-1}$ if its least period is $2^{k}$. The upper bound follows from the fact that the linear complexity of a sequence equals its period if and only if such a sequence is $'0000 \ldots 01'$. Nevertheless, $'0000 \ldots 01'$ is neither a balanced sequence nor an element of the generalized self-shrinking sequence Abelian group.
\hfill $\Box$

A more refined result can be stated for the self-shrinking sequence.
\begin{corollary}\label{corollary:4}
The linear complexity of a self-shrinking sequence with least period $T = 2^{L-1}$ satisfies
\begin{equation}\label{equation:18}
2^{L-2} < LC < 2^{L-1} - (L - 2).
\end{equation}
\end{corollary}
\textit{Proof:} The lower bound is the same as that of (\ref{equation:17}) whereas the upper bound was proved in \cite{Blackburn:1999}.
\hfill $\Box$

Remark that the lower bound for the linear complexity of the self-shrinking sequence given by Meier and Staffelbach in [6]
\begin{equation}\label{equation:19}
LC_{ss} \geq 2^{\lfloor L/2 \rfloor -1},
\end{equation}
is much less than the lower bound given in (\ref{equation:18}).

\subsection{Pseudorandomness in the generalized self-shrinking sequences}
Until now, we have found that the generalized self-shrinking sequences exhibit good cryptographic properties: long period, large linear complexity, balancedness, excellent correlation, etc.
So the question remains whether or not this sequence family can be used for stream cipher.
In terms of mappings, the runs of several generalized self-shrinking sequences are now analyzed. More precisely, we enhance different elements of this family that never must be used for cryptographic purposes.
\begin{lemma}\label{lemma:5}
Denote $2 p = q$ with $p$ defined in lemma \ref{lemma:1}. Let $\alpha^q$ and $\alpha^{q+1}$ be elements of $GF(2^L)$. Then, the following equality holds
\begin{equation}\label{equation:20}
\alpha^{q} + \alpha^{q+1} = \alpha^p .
\end{equation}
\end{lemma}
\textit{Proof:} From (\ref{equation:6}), we get $\alpha^{q+2} = \alpha^{q} + 1$. Thus,
\begin{eqnarray}\label{equation:21}
\alpha^{q} + \alpha^{q+1} &=& (\alpha^{q+2} + 1) + \alpha^{q+1} =\alpha^{q+1} (\alpha + 1) + 1 \nonumber \\
&=& \alpha^{q+1} \cdot \alpha^{-p} + 1= \alpha^{p+1} + 1  =  \alpha^{p} .
\end{eqnarray}
\hfill $\Box$
\begin{lemma}\label{lemma:6}
According to lemma \ref{lemma:5}, the following expressions can be written
\begin{eqnarray}\label{equation:22}
\alpha^{q} &=& c_{L-1} \, \alpha^{L-1} + \ldots + c_1 \alpha + c_0 \nonumber\\
\alpha^{q+1} &=& c'_{L-1} \, \alpha^{L-1} + \ldots + c'_1 \alpha + c'_0 \\
\alpha^{q+2} &=& c_{L-1} \, \alpha^{L-1} + \ldots + c_1 \alpha + \bar{c}_0 \nonumber,
\end{eqnarray}
where $c_j, c'_j \in GF(2)$. For these powers of $\alpha$,
if $c_{L-1} = 0$, then $c'_{L-1} = 1$ and viceversa.
\end{lemma}
\textit{Proof:}  Both cases are to be considered.
\begin{enumerate}
\item For $c_{L-1} = 0$: let $i$ be the greatest index for which $c_i \neq 0$. Then
\begin{eqnarray}\label{equation:23}
\alpha^{q} &=& c_{i} \alpha^{i} + c_{i-1} \alpha^{i-1}+\ldots + c_0 \nonumber \\
\alpha^{q+1} &=& c_{i} \alpha^{i+1} + c_{i-1} \alpha^{i}+\ldots +  c_0 \alpha \\
\alpha^{q+2} &=& c_{i} \alpha^{i+2} + c_{i-1} \alpha^{i+1}+\ldots +  c_0 \alpha^2 \nonumber.
\end{eqnarray}
The group of equations (\ref{equation:22}) and (\ref{equation:23}) hold simultaneously only for $i = L-2$. That is, $c_i = c_{L-2} = 1$. Therefore,
\begin{eqnarray}\label{equation:xx}
\alpha^{q} &=& 0 \cdot \alpha^{L-1} + 1 \cdot \alpha^{L-2} +\ldots + c_0 \nonumber\\
\alpha^{q+1} &=& 1 \cdot \alpha^{L-1} + c_{L-3} \alpha^{L-2} +\ldots + c_0 \alpha. \nonumber
\end{eqnarray}
Thus, $c'_{L-1} = 1$ and the power $\alpha^{L-1}$ is included in the linear decomposition of $\alpha^{q+1}$.
\item For $c_{L-1} = 1$:
\begin{eqnarray}\label{equation:yy}
\alpha^{q} &=& \; 1 \cdot \alpha^{L-1} + c_{L-2} \alpha^{L-2} +\ldots + c_0 \nonumber\\
\alpha^{q+1} &=& c'_{L-1} \alpha^{L-1} + c'_{L-2} \alpha^{L-2} +\ldots + c'_0 \alpha. \nonumber
\end{eqnarray}
From lemma \ref{lemma:5},
\begin{eqnarray}\label{equation:24}
\alpha^{q} + \alpha^{q+1} &=& \\
(1 + c'_{L-1}) \alpha^{L-1} + \ldots + (c_1 + c'_0) \alpha + c_0 &=& \alpha^p. \nonumber
\end{eqnarray}
According to corollary (\ref{corollary:1}), $\alpha^p$ includes the power $\alpha^{L-1}$ in its linear decomposition. Thus, $c'_{L-1} = 0$.
\end{enumerate}
\hfill $\Box$

\begin{theorem}\label{theorem:6}
The mapping  of $1 \rightarrow \alpha^{q+1}$ generates a generalized self-shrinking sequence where all its runs (blocks and gaps) have even length.
\end{theorem}
\textit{Proof:} Writing the mapping $1 \rightarrow \alpha^{q+1}$, we have:
\[
\begin{tabular}{l}
$1 \;\,\,\rightarrow \alpha^{q+1}$\\
$\alpha \,\;\rightarrow \alpha^{q} + 1$\\
$\alpha^2 \rightarrow \alpha^{q+1} + \alpha$\\
$\alpha^3 \rightarrow \alpha^{q} + 1 + \alpha^2$\\
$\alpha^4 \rightarrow \alpha^{q+1} + \alpha + \alpha^3$\\

\;\vdots \qquad \qquad \quad \vdots\\
$\alpha^{m} \rightarrow \alpha^{q+1} + \alpha + \alpha^{3} + \ldots + \alpha^{m-1}$ (if $m$ is even) or\\
$\alpha^{m} \rightarrow \alpha^{q} \;\;\;\, + 1 \, + \alpha^{2} + \ldots + \alpha^{m-1}$ (if $m$ is odd).
\end{tabular}
\]
Let $\alpha^{n_k}$ $(0 \leq k < 2^{L-1})$ be the $2^{L-1}$ powers of $\alpha$ that include $\alpha^{L-1}$ in their linear decomposition, that is $c_{L-1} \neq 0$. We add pairs of consecutive powers $(\alpha^{n_i}, \alpha^{n_{i+1}})$ with $i = \dot{2}$, $(i = 0, 2, 4, 6, \ldots)$.

The key idea is to prove that $\alpha^{n_i} + \alpha^{n_{i+1}}$ includes the power $\alpha^{L-1}$ an even number of times. In that case, $v_{n_i} + v_{n_{i+1}} =0$ in the sliding sequence and every two consecutive terms in the generalized self-shrinking sequence will be equal. In fact, given the pair $(\alpha^{n_i}, \alpha^{n_{i+1}})$ different cases may occur:
\begin{enumerate}
\item $(n_i, n_{i+1})$ are integers (even, odd) or (odd, even), then
\begin{equation}\label{equation:25}
\alpha^{n_i} + \alpha^{n_{i+1}} = \alpha^{q} + \alpha^{q+1} + \sum _{k=0}^{n_i} \alpha^k.
\end{equation}
The number of powers $\alpha^{L-1}$ in (\ref{equation:25}) is:
\begin{itemize}
\item Once in $\alpha^{q}$ or in $\alpha^{q+1}$ by lemma \ref{lemma:6}.
\item $i$ times from the pairs $(\alpha^{2j}, \alpha^{2j+1})$ with $(j = 0, 1, \ldots, i/2 - 1)$.
\item Once in $\alpha^{n_i}$.
\end{itemize}
Thus, the number of powers $\alpha^{L-1}$ is always even.

\item $(n_i, n_{i+1})$ are integers both even or both odd, then
\begin{equation}\label{equation:26}
\alpha^{n_i} + \alpha^{n_{i+1}} = \sum _{k} \alpha^k \qquad n_i < k < n_{i+1},
\end{equation}
where the summation is extended to powers of $\alpha$ not including the term $\alpha^{L-1}$.
\end{enumerate}
Thus, this generalized self-shrinking sequence only exhibits runs of even length.
\hfill $\Box$

Next a different generalized self-shrinking sequence with a wrong run distribution is introduced.
\begin{theorem}\label{theorem:7}
The mapping  of $1 \rightarrow \alpha^{q}$ generates a generalized self-shrinking sequence where all its runs (blocks and gaps) have length 1 or 2 exclusively.
\end{theorem}
\textit{Proof:} From (\ref{equation:20}) $\alpha^{q} = \alpha^{q+1} + \alpha^p$. Moreover, by theorem \ref{theorem:6} and corollary \ref{corollary:1} the generalized self-shrinking sequence associated to the mapping $1 \rightarrow \alpha^{q}$ is the bit-wise addition of a sequence with runs of even length and the sequence $'1010 \ldots 10'$. Then, for $u, \bar{u} \in GF(2)$ we have
\begin{itemize}
\item A run of even length and $'1010 \ldots 10'$
\[
\begin{tabular}{c}
$u \, u \, u \, u \,  \ldots u \, u  + 1010 \ldots10  =  \bar{u}\, u\, \bar{u}\, u\, \ldots \bar{u}\, u $\\
\end{tabular}
\]
generates runs of length $l = 1$.
\item The succession of a block and a gap (or viceversa)
\[
\begin{tabular}{c}
$\ldots \bar{u} \, \bar{u} \, u \, u  + \ldots 1010  =  \ldots u\, \bar{u}\, \bar{u}\, u$
\end{tabular}
\]
generates a run of length $l = 2$.
\end{itemize} \hfill $\Box$

Complemented sequences of these sequences in theorems \ref{theorem:6} and \ref{theorem:7} also exhibit undesirable runs for cryptographic purposes.

\begin{table}[!t]
\renewcommand{\arraystretch}{1.3}
\caption{Nonpseudorandom generalized self-shrinking sequences}
\label{Table:4}
\centering
\begin{tabular}{|l||c|}
\hline
Mapping: $1 \rightarrow \alpha^{s}$ & $\{b_n(\alpha^s)\}$\\
\hline \hline
$1 \rightarrow \alpha^{q+1}$ & $0000110011001111$\\
\hline
$1 \rightarrow \alpha^{q+1} + 1$ & $1111001100110000$\\
\hline \hline
$1 \rightarrow \alpha^{q}$ & $1010011001100101$\\
\hline
$1 \rightarrow \alpha^{q} + 1$ & $0101100110011010$\\
\hline
\end{tabular}
\end{table}

Table \ref{Table:4} shows the generalized self-shrinking sequences corresponding to the mappings $1 \rightarrow \alpha^{q+1}$ and $1 \rightarrow \alpha^{q}$ as well as their complemented sequences ($1 \rightarrow \alpha^{q+1} + 1$ and $1 \rightarrow \alpha^{q} + 1$) for $L = 5$ and $p(x) = x^5 + x^2 + 1$ with $q = 26$.

\section{Conclusions}
This work tackles the problem of the least periods in the family of generalized self-shrinking sequences. Moreover, based on a formulation of linear mappings the conjecture formulated by Hu and Xiao in \cite{Hu:2004} has been proved. As a straight consequence of the result, other cryptographic parameter, linear complexity, has been perfectly quantified. In addition, the period of the self-shrinking sequence, an element of this family, is given as well as its linear complexity is lower and upper bounded. The values of both parameters improve dramatically the results given by Meier and Staffelbach in \cite{Meier:1995}.

At first glance, the generalized self-shrinking sequences seem to satisfy good cryptographic properties: long period, large linear complexity, balancedness, excellent correlation, etc. Nevertheless,
concerning the pseudorandomness of this family, it is showed that although some sequences exhibit good pseudorandomness e.g., the self-shrinking sequence as reported in \cite{Meier:1995}, there are other ones whose distribution and run length make them undesirable for cryptographic purposes. In this sense, the elements of this family should be carefully analyzed before to be recommended for their use in stream cipher.

\end{document}